\documentstyle[12pt,aasms4,epsfig,rotate]{article}
\newcommand{\etal}{ {\it et al.}}

\begin{document}

\title{Broad band high energy observations of the superluminal jet source GRO 
J1655-40 during an outburst}
\author{S. N. Zhang$^{1,2}$, K. Ebisawa$^{2,3}$, R. Sunyaev$^{4,5}$, 
Y. Ueda$^{6}$, B. A. Harmon$^{1}$, S. Sazonov$^{4,5}$, G.~J.~Fishman$^{1}$,
H. Inoue$^{6}$, W. S. Paciesas$^{1,7}$, T. Takahashi$^{6}$}
\affil{$^{1}$ES-84, NASA/Marshall Space Flight Center, Huntsville, AL 35812, USA\\
$^{2}$Universities Space Research Association\\
$^{3}$Code 660.2, NASA/Goddard Space Flight Center, Greenbelt, MD 20771, USA\\
$^{4}$Space Research Institute, Moscow, Russia\\
$^{5}$Max-Planck-Institute f\"{u}r Astrophysik, Garching, Germany\\
$^{6}$Institute of Space and Astronautical Science, Kanagawa, Japan\\
$^{7}$University of Alabama in Huntsville, Huntsville, Alabama\\
}

To appear on the April 10, 1997 Issue of ApJ, Part I

\abstract{
The X-ray/radio transient superluminal jet source\\ GRO~J1655-40 was recently
suggested to contain a black hole from optical observations. Being a relatively
close-by system (d$\sim$3.2 kpc), it can likely provide us with rich
information about the physics operating in both galactic and extragalactic jet
sources. We present the first simultaneous broad band high energy observations
of GRO~J1655-40 during the 1995 July-August outburst by three instruments: ASCA, WATCH/GRANAT and
BATSE/CGRO, in the energy band from 1 keV to 2 MeV. Our observations strengthen
the interpretation that GRO~J1655-40 contains a black hole. We detected a
two-component energy spectrum, commonly seen from other galactic black hole
binaries, but never detected from a neutron star system. Combining 
our results with the
mass limits derived from optical radial velocity and orbital period
measurements, we further constrain the mass of the central object to be between
3.3 and 5.8 M$_{\sun}$, above the well-established mass upper limit of 3.2
M$_{\sun}$ for a neutron star (the optical mass function for GRO~J1655-40 is
3.16$\pm$0.2 M$_{\sun}$). This system is therefore the first galactic
superluminal jet source for which there is strong evidence that the system
contains a stellar mass black hole. The inclination angle of the binary system
is constrained to be between 76 and 87 degrees, consistent with estimates
obtained from optical light curves and radio jet kinematics.} 


\section{Introduction}

Radio observations are now beginning to show that relativistic jets may be a
more common feature of X-ray binary systems than previously thought.
Some, such as
SS433 
(Margon 1988), Cygnus~X-3 
(Strom \etal ~   1989) 
and the more recently
discovered superluminal transients GRS~1915+105 
 (Mirabel and Rodriguez 1994) 
and GRO~J1655-40 
 (Tingay \etal ~   1995; Hjellming and Rupen 1995) 
show
proper motions implying velocities V$\sim$0.2-0.9 C. Of those, only
GRO~J1655-40 has a compact object whose dynamically estimated mass is greater
than 3 M$_{\sun}$ 
(Bailyn \etal ~    1995b). 
{\it It is thus generally believed to be
a black hole binary} (BHB). Being a BHB with relativistic jet ejection episodes,
GRO~J1655-40 is a good analogue to the active galactic nuclei (AGN), which are
thought to contain super-massive black holes and frequently are observed to have
relativistic jets associated with the central engine. 

The galactic jet sources are of great interest because the dynamic time scales
of BHBs are a factor of $\sim$10$^{6}$-10$^{9}$ smaller than those of typical
AGN 
(Shakura \& Sunyaev 1976). 
Therefore dynamical variations of AGN on time scales from many years up to
the Hubble time correspond in BHBs to intervals of seconds
to years. Observations of jet formations and decay in concidence with studies 
overall flux variations and spectral evolution allow the possibility of 
identifing the emission regions at various wavelengths.

Studying the inner part of the accretion disk very close to the central black
hole is critical for understanding the central engine and the
jet ejection. This region can be observed at X-ray and hard X-ray wavelengths
for BHBs. For AGN, however, this region is much more difficult to observe. This is because
the typical photon energy radiated from the inner accretion disk region of an
AGN is much lower ($\sim$eV). Absorption of such low energy photons in the
complex surrounding environment and the interstellar medium is severe. Thus
GRO~J1655--40 {\it is a unique system for understanding the physics involved 
in accreting black holes and relativistic jets in our own Galaxy and AGN.} 

Since the discovery of GRO~J1655--40 
(Zhang \etal ~    1994a) 
by the Burst and
Transient Source Experiment (BATSE) aboard the NASA Compton Gamma-Ray
Observatory (CGRO) on July 27, 1994, it has been extensively observed in radio,
optical, X-ray and gamma-ray bands. Radio observations have revealed several
relativistic ejection episodes 
(Tingay \etal ~   1995; Hjellming \& Rupen 1995) 
from this system,
following hard X-ray (above 20 keV) activity 
(Harmon \etal ~    1995). 
Optical
observations 
 (Bailyn \etal ~    1995a; 1995b) 
have determined the orbital period
(2.62 days), the nature of the companion star (F-type) and the mass limit of the
central compact object ($\ge$3 M$_{\sun}$). 

The BATSE experiment detected 4 major outbursts from this system separated by
about 120 days 
(Zhang \etal ~    1995a). 
Two of them were also observed
by the WATCH instrument on the GRANAT satellite 
(Sazonov, Sunyaev \& Lund 1996). 
The energy
spectrum above 20 keV when the source is in outburst is well described by a
power-law (photon index 1.5-3.1) up to at least 200 keV. OSSE/CGRO detected
emission up to 600 keV without any detectable deviation from the 
power-law 
(Kroeger \etal ~   1996). 
Searches in both BATSE and OSSE data for rapid
variability resulted in upper limits of about 5\% of integrated r.m.s.
noise from 0.01-1 Hz 
(Kroeger \etal ~   1996; Crary \etal ~   1996). 
In light of
the possible optical eclipses 
(Bailyn \etal ~   1995a; but cf. van der Hooft \etal ~   1996), 
both BATSE and OSSE data were searched for orbital
eclipses and/or modulation. No indications of periodic
behavior were found 
(Zhang, \etal ~ 1996a; Kroeger \etal ~  1996). 

The TTM
instrument (2-27 keV) aboard the Russian MIR space station also observed this
source on several occasions (Sept. and Oct. 1994 and Feb. 1995) between the
main hard X-ray outbursts detected by BATSE. An ultra-soft spectrum, compatible
with an accretion disk origin, was observed during most of these 
observations 
(Alexandrovich \etal ~  1994; 1995).
Sometimes a hard X-ray tail was seen by TTM and HEXE (15-200 keV) also aboard
the MIR station 
(Sunyaev \etal ~ 1996). 
X-ray observations of GRO~J1655-40 in the 1-10 keV range with the Japanese ASCA satellite have
revealed very unusual spectral characteristics. On two occasions, Aug. 23 and 
Sep. 27, 1994, when the
source flux in the BATSE energy band (20-100 keV) was very low ($<$30-50
mCrab), ASCA detected an energy spectrum with a significant high energy
cutoff above 4-5 keV. Absorption line features
were seen in the region of the Fe K-edge 
(Ueda \etal ~ 1996). 

\section{Multi-instrument observations}

The outburst in July-August 1995 was simultaneously observed from 1 keV to 2
MeV, by ASCA (1--10 keV), WATCH/GRANAT (8-20 keV) and
BATSE/CGRO (20-2000 keV) instruments. The X-ray telescope 
 (Tanaka, Inoue \& Holt, 1994) 
aboard a Japanese satellite ASCA uses focusing X-ray optics to
concentrate X-ray photons in the 1-10 keV range onto its detector planes, the
solid state spectrometer (SIS) and the gas imaging spectrometer (GIS). The SIS
was not usable for this observation due to the brightess of the source
which caused telemetry saturation and
pulse pile-up. The GIS was usable for energy spectral measurements, though
significant deadtime was present. Deadtime was corrected for in estimating the
incident source flux. The ASCA observation occurred between Aug. 15.45-16.18
(UT), during the rise to the second peak of this outburst. Overall flux
variations during the nearly one-day observation were less than 10\%.

The WATCH instrument 
(Lund 1986) 
aboard the Russian GRANAT spacecraft detected this
outburst while the GRANAT Observatory was scanning the sky. 
In the scanning mode the motors of the WATCH collimators do
not rotate, and the spinning of the satellite is used to produce modulation
patterns. A detailed analysis of the data was done 
to separate the contributions of GRO~J1655-40 and the nearby source 4U~1700-37
to the modulation curve. Observations during the X-ray eclipses of 4U~170037 and 
simulations of the instruments's imaging capability gave proof that the
obtained light curve of GRO~J1655-40 is correct. 

GRO~J1655-40 has been continuously monitored by BATSE's Large Area Detectors
(LADs) 
(Fishman \etal ~   1989) 
using Earth occultation analysis 
(Harmon \etal ~    1992) 
and Earth occultation transform imaging 
(Zhang \etal ~ 1993; 1994b; 1995b) 
techniques. The un-collimated nature of the LADs
makes other active sources also detectable during these outbursts. The data are
thus simultaneously fitted with these sources included to determine the net
contributions from GRO~J1655--40. 

\section{Broad energy band spectrum}

The ASCA-BATSE measured spectrum is shown in figure 1. The ASCA data 
cannot be fit by standard single-component models. However, they are
consistent with a two-component spectral model, consisting of a power-law
component and a multi-color blackbody disk (MCBD) component 
(Shakura \& Sunyaev 1973 and see Mitsuda \etal ~ 1984, Makishima \etal ~   1986 for the MCBD model).
The flux detected by WATCH in the 8-20 keV during
this time was 620$\pm$50 mCrab (here the Crab 
flux was determined contemporaneously with the GRO~J1655-40 observation).
The WATCH
detection is also plotted in the same figure. Extrapolation of the BATSE
power-law spectrum into the WATCH energy range gives $\sim$240 mCrab. The WATCH
flux thus independently suggests the presence of some additional soft component, which
should contribute $\sim$380 mCrab in the 8-20 keV flux. The ultra-soft
component observed with ASCA fitted by the MCBD model would produce $\sim$400
mCrab in the WATCH energy band; therefore the value of the flux measured with WATCH
is in a good agreement with a model consisting of a power-law component and a
MCBD component. No other spectral component is needed to account for the 1-200
keV overall spectrum, except for a weak absorption feature in the Fe
K-fluorescence energy
band of the ASCA data 
(Ueda \etal ~  1996). 
The ASCA continuum model parameters
for this two-component spectrum are: neutral hydrogen column density N$_{\rm
H}$=(8.9$\pm$0.3)$\times$10$^{21}$~cm$^{-2}$, inner disk edge temperature
kT$_{\rm in}$=1.36$\pm$0.01 keV, inner disk radius R$_{\rm
in}\sqrt{cos(i)}$=9.3$\pm$0.2 km (@3.2 kpc), where $i$ is the inclination angle
of the system, power law photon spectral index $\alpha$=2.36$\pm$0.08 and the
normalization factor a$_{0}$=6.3$\pm0.9$ $ph \cdot s^{-1} \cdot cm^{-2} \cdot
keV^{-1}$ at 1 keV for the power-law. The independently determined BATSE power-law parameters are:
photon spectral index $\alpha = 2.43 \pm 0.3$ and flux $1.05 \times 10^{-3} \pm
1.7 \times 10^{-4} ph \cdot s^{-1} \cdot cm^{-2} \cdot keV^{-1}$ at 40 keV.
The ASCA and BATSE power-laws are consistent with each other within their
statistical errors. 

\placefigure{figure1}

\section{Source intensity variability}

Source intensity variations on time scales of seconds to hundreds of seconds
can be studied with both ASCA and BATSE. Telemetry saturation and deadtime
make it impossible to study short-term variations using the ASCA GIS event data
which we used to obtain the energy spectrum discussed in the last section. 
However, the
GIS monitor counter (type `L1'), which has neither position nor energy
resolution is usable for short-term variation studies.
All the GIS events which passed the pulse-height (0.9 to 10 keV) and rise-time
discriminators are recorded as monitor counts with a time resolution of 0.125
sec. In figure 2, we show a power density spectrum (PDS) of GRO~J1655--40
calculated from the GIS monitor counts in the 0.002 -- 4 Hz band, together with
those of two black hole candidates (BHCs), Cyg~X-1 in a low state and 
the X-ray nova GRS~1009--45 in a high state 
(Tanaka 1994; Sunyaev et al.\ 1994), 
for comparison.  The Poisson noise level has been subtracted.
The GRO~J1655--40 PDS clearly shows excess power above the Poisson noise level,
and is roughly represented by a power-law, $f^{-1.2}$.  The r.m.s. variation 
relative to the average flux is $5.5\pm0.2$\%. For comparison the r.m.s.
fraction for Cyg~X-1 is $31.4\pm0.3$\%. Significant short-term variations
were not found from GRS~1009--45 in excess of the Poisson noise level. 

\placefigure{figure2}

The continuous counting rates of the BATSE LADs at a time resolution of 1.024
seconds were also analysed for evidence of variability. The 20-100 keV flux, however,
showed no variation at an upper limit of about 20\% integrated from 0.01 to
0.488 Hz (Crary, D., private communication, 1996). A smaller upper limit
($\sim$5\%) was obtained for other brighter outbursts from this source 
(Crary \etal ~    1996). 

\section{Light curves and spectral variations}

In figure 3 we show the WATCH (8-20 keV) and BATSE (20-100 keV) light curves
and spectral variations during the July -- August 1995 outburst. The double peak structure and overall spectral
evolution are similar to that of the previous three major outbursts from this
source 
 (Zhang \etal ~  1996a). 
It is interesting to compare
the BATSE and WATCH measurements throughout the entire outburst. In Figure 3
(upper panel), we also plot the extrapolated BATSE flux in the 8-20 keV band
from the measured fluxes in 20-100 keV band,
together with the WATCH light curve. The extrapolated BATSE 8-20 keV light curve should be
dominated by the power-law component since any thermal component from the
accretion disk cannot contribute to the 20-100 keV band power-law
significantly. It is clear that  whenever the source was simultaneously detected
by both WATCH and BATSE, the WATCH flux level is almost always significantly
above the power-law component projected from BATSE measurements. 
The ASCA detection of the disk blackbody component suggests
that all the additional flux observed by WATCH originates from the accretion 
disk. It is
interesting to note that during TJD 9947-9956, the 8-20 keV flux is dominated by the
thermal component, while the 20-100 keV power-law is very hard (photon spectral
index $\sim$1.5-2.0). 

\placefigure{figure3}

\section{Discussion}

\subsection{GRO~J1655--40 as a black hole system}

For the first time we have detected a two-component energy spectrum from 
a galactic superluminal jet source during a hard X-ray outburst.
The thermal component fits well with the MCBD
model commonly used to describe the X-ray emission from an optically thick and
geometrically thin accretion disk. Using the measured energy spectrum, we can
further constrain the parameters of the binary system. From the MCBD model 
fitting,
the inner disk radius is constrained to be R$_{\rm in}\sqrt{cos(i)}$=9.3$\pm$0.2 km
(assuming the source is at a distance of 3.2 kpc). From the relationship
between the Schwarzschild radius R$_{\rm s}$ and the inner disk radius
R$_{\rm in}$, 3R$_{\rm s}\approx (3/5) \times 0.8 \times {\rm R}_{\rm in}
\times f^{2}$, we can obtain the relationship between the mass of the central
compact object ($M$) and the inclination angle of the system ($i$). In the
above equation, the factor 3/5 accounts for the disk effective temperature
becoming a maximum at $\sim$5 R$_{\rm s}$, and 0.8 is the relativistic
correction for gravity becoming effectively stronger than in the Newtonian
case. The factor $f$ is the ratio between the color temperature and the
effective temperature and should be $\sim$1.7 (see 
Shimura \& Takahara, 1995
for details). 

In figure 4, we plot the primary mass vs inclination ({\it M--i}) 
derived from optical (mass
function of 3.16$\pm$0.2 M$_{\sun}$ 
(Bailyn, \etal ~   1995b)) 
and X-ray
spectroscopic observations. For the X-ray {\it M--i} relationship, we also plot the
lower and upper bounds derived from the 3-$\sigma$ error of 0.6 km for R$_{\rm
in}\sqrt{cos(i)}$ combined with the lower and upper limits of 3.0 and 3.5 kpc
for the source distance 
(Hjellming \& Rupen 1995). 
A larger distance shifts the curve upward. The allowed
parameter space is shown on the figure as the shaded area. We find the mass of
the compact object lies between 3.3 and 5.8 M$_{\sun}$ for all possible
parameters. {\it This lower mass limit is solidly above the well established neutron
star mass upper limit.} The range of the inclination angle is between 76 and 87
degrees, consistent with the optical light curve modulation 
(Bailyn \etal ~   1995a, 1995b; van der Hooft \etal ~   1996) and 
radio jet kinematics solutions 
(Hjellming \& Rupen, 1995). 
We therefore conclude that {\it this superluminal jet source contains a
black hole and has the highest inclination angle of all galactic BHBs
with well determined system parameters.} 

\placefigure{figure4}

The allowed parameter space can be further divided into two areas, labeled as
{\bf `A'} and {\bf `B'}. For a minimum possible companion mass of 0.1
M$_{\sun}$ (below which such a binary system would not be formed), a
sufficiently high inclination angle would result in the center of the accretion
disk to be eclipsed by the companion. This `eclipse' line is depicted by the
nearly vertical thin {\it solid} line. The parameter space on the right side of
this line is a part of the {\it eclipsing zone}. For a lower inclination angle,
a more massive companion will also result in such eclipse. The `eclipse' line
corresponding to 0.32 M$_{\sun}$ is shown as the vertical thin {\it dotted}
line. The 0.32 M$_{\sun}$ is chosen because this line intersects with the
optical {\it M--i} relationship for the same companion mass on the upper bound of
the X-ray {\it M--i}. Therefore if the system parameters fall in the area {\bf `B'},
the accretion disk center will be eclipsed by the companion. In area {\bf
`A'}, no such eclipse will happen. The mass of the companion cannot be higher
than 0.32 M$_{\sun}$ for a non-eclipsing system.

A reasonable mass of the companion is between 0.5-1.5 M$_{\sun}$. In this case,
this system will be the first detected eclipsing BHB in our Galaxy. Eclipse
mapping may be applied to determine the details of the inner disk region for
the first time. The BH mass lies between 4.1 and 5.3 M$_{\sun}$ and the
inclination angle is between 80 and 86.5 degrees. The inner disk radius can be
estimated to be between 25 and 36 km, considerably larger than that of neutron
star binaries ($\sim$10 km). On the other hand, this system has an unusually
large space velocity, indicating that it probably has experienced an unusual
evolution 
(Brandt, Podsiadlowski \& Sigurdsson, 1995). 
Therefore the companion
could possibly be severely under-massive. Constraints from evolutionary theories
suggest that the companion cannot be less massive than 0.23 M$_{\sun}$ 
 (Brandt, Podsiadlowski \& Sigurdsson, 1995). 
From figure~4 it is clear that the allowed
parameter space between 0.23-0.5 M$_{\sun}$ companion is very small. This
system may also be an important object for our understanding of the formation
and evolution of BHBs. 

A further support that the system contains a black hole is the lack of an
additional thermal component commonly seen from a neutron star surface with a
higher temperature (typically a $\sim$2 keV blackbody spectrum, which
is considerably harder than the MCBD spectrum, i.e., the
ultra-soft component). The ultra-soft component by itself, however, cannot be
taken as a signature of a black hole, since 4U~0142+614, a neutron star 
pulsar system,
has displayed such an ultra-soft spectrum 
(Mereghetti, Stella \& De Nile 1993; Israel, Mereghetti \& Stella 1993). 
Because of the mass
accretion rate dependency of the disk temperature, a neutron star system may be
observed with a very low disk temperature when its mass accretion rate is low
and therefore the X-ray luminosity is low. We therefore argue that {\it a luminous
($\sim1 L_{\rm Eddington}$ for 1 M$_{\sun}$) ultra-soft X-ray spectrum is a
very strong indication of a black hole system} and should be used for selecting
BHCs. A prominent power-law component associated with such a bright ultra-soft
component, however, has never been observed from a neutron star system but is
now seen from persistent, transient and jet type BHBs. {\it Such a 
two-component spectrum
may, therefore, be
a firm signature from a black hole system} 
(Sunyaev \etal ~  1988, 1994; van der Klis 1993; van der Klis \&
van Paradijs 1993; White, 1993; Tanaka \& Lewin 1995; Zhang \etal ~   1996b). 

\subsection{Origin of the power-law hard X-ray tail}

We are reasonably confident that the ultra-soft component originates from the
accretion disk due to the gravitational energy release of the accreted material
in the X-ray energy band. The origin of the hard X-ray power-law component is,
however, more controversial, although inverse Compton up-scattering of low
energy photons by fast moving particles (electrons or protons) is commonly
accepted as the radiation mechanism. The non-detection of short-term variations
from the hard X-ray power-law component suggests a different origin of this
power-law than for the low state hard X-ray power-laws (with cutoff) observed from both black
hole and neutron star binaries. Thermal Comptonization models normally used to
interpret the low state hard X-ray power-laws cannot be applied here,
regardless of the nature and geometries of the hot Compton cloud,
because
the relatively steep power-law extended to a very high energy implies a very high
electron temperature (50-200 keV) and a very small optical depth (0.01-0.1). It
is very hard to keep this optically thin cloud sufficiently stable to account for the low
(or no) short-term variability in the hard X-ray power-law component. Cooling
of the high energy electrons by the copious soft X-ray photons in the high
state seems to be an additional complication, unless there is a powerful underlying
heating mechanism. 

Based on the detected relativistic jets from this source and the correlation of
radio flares with  hard X-ray activity, it has been suggested that the fast
moving particles responsible for the radio emission from the jets 
 (Levinson \& Blandford 1996) 
also scatter low energy photons via the inverse Compton process
to produce the observed hard X-rays. We believe this is unlikely since this and
the previous outburst were not detected to have any radio emission, unlike the
first two outbursts with strong radio emission. The overall radio emission
decayed exponentially since the initial hard X-ray outburst, while the hard
X-ray outbursts with similar peak luminosities occurred four times since the initial outburst with a
separation of about 120 days. Absence of the jet-ejection and any radio
emission at all during the last two hard X-ray outbursts argues strongly
against the scenario where hard X-rays are produced directly or indirectly from
the jets. 

The upper limits of about 0.5 mJy at 3.6 cm (Hjellming 1996) 
allows us to set a limit of the size of the possible radio emission region.
The surface brightness (due to incoherent synchrotron radiation) temperature 
upper limit of 10$^{12}$ k (Shu 1991) indicates that the dimension of the possible radio
emission region is less than 1.7$\times$10$^{11}$ cm (Hjellming, R.M, private
communication). This is smaller than the size of the binary system 
($\sim$10$^{12}$ cm), but can still be several orders of magnitude larger than
the soft X-ray (blackbody) emission region ($\sim$10$^{7}$ cm). 
Since this region is so close to and very likely covers completely 
the soft photon emission region, the high energy electrons
required for producing synchrotron radiations will be cooled down effectively
by the inverse Compton scattering process,
thus preventing significant synchrotron radiations. Therefore the power law hard X-ray
photons cannot be produced via the synchrotron radiation mechanism.

Another model involves the converging flow of fast moving
material near the black hole horizon 
(Chakrabarti \& Titarchuk 1995). 
This
model does not have the difficulties discussed above for other models. In this
model, the power-law component is produced via the inverse Compton scattering
of low energy photons of the ultra-soft component by the bulk motion of the
converging flow. This model predicts a spectral break energy of $\sim$511 keV
divided by the total mass accretion rate in terms of the Eddington rate
corresponding to the central compact object mass. The detected luminosity
during the ASCA observation is $\sim$ 10$^{38}$ erg/s and could be a factor of
2-5 higher at other times (scaled up from the detected WATCH and BATSE fluxes).
Therefore the Eddington accretion rate is $\le$ 1 for a 3-5
M$_{\sun}$ black hole. The unbroken power-law up to 200-600 keV is certainly
consistent with this prediction. It is also interesting to note that a
relativistic converging flow cannot take place near a neutron star surface when
the mass accretion is high due to the radiation pressure from the neutron star
surface. {\it This provides a possible explanation why such a two-component 
spectrum has never been observed from a neutron star binary.} We consider, 
however, that it is
still difficult within this model to explain the range of spectral
variations during this outburst, expecially the very hard power-law with a huge
thermal excess at low energies near the end of this outburst. If GRO~J1655-40
were indeed an eclipsing system, this model would be excluded since it requires the
converging flow and thus the hard X-ray production region to be very close to the BH.
However, no hard X-ray eclipse or any significant hard X-ray orbital
modulation is seen in the BATSE or OSSE data. 

\subsection{Origin of the short-term intensity variations}

Short-term variability was detected for the first time from a galactic
superluminal jet source in a high state. In view of the low (or no) variability in
the 20-100 keV band, there are at least two possibilities to account for the
detected short-term variability in the 1-10 keV band. One is that the
ultra-soft component is variable. We consider this to be unlikely since no similar
variability is seen from the ultra-soft component of other BHBs, unless the
nature of this ultra-soft component is different due to the jet-ejection nature
of this system. (However, the energy spectrum does not look different from other
BHBs.) Two other BHCs, namely GX339-4 
(Miyamoto \etal ~   1991) 
and GS~1124-68 
(Miyamoto \etal ~   1993, 1994; Ebisawa \etal ~   1994) 
in their high state have been
detected with a combination of strong hard X-ray power-law tails and moderate
($\sim$5\%) r.m.s. variations. These variations were found to be related to the
power-law tails and such a power-law-like PDS from GRO~J1655-40 was not seen from
them. 

The other possibility is that the power-law component is variable, but the
variability has an energy dependency so that the 1-10 keV power-law is
significantly variable but the 20-100 keV power-law is not. To account for the
5.5\% variability with only about 25\% of the 1-10 keV flux in the power law
component, the intrinsic variability in the 1-10 keV power-law component has to
be as high as 20\%. Energy dependent variations have been observed previously
from other BHBs 
 (Ebisawa, \etal ~   1994, Miyamoto \etal ~   1991). 
The combined
1-10 keV variability and 20-100 keV low (or no) variability is, however,
unusual among BHBs. 

Between the hard X-ray outbursts, almost identical short-term variability has 
also been observed by ASCA from GRO~J1655-40 and GRS~1915+105 
(Ebisawa \etal ~  1996). 
GRS~1915+105 is the only other
detected galactic superluminal jet source. Its energy spectrum
(power-law with cutoff above 5 keV, plus {\it Fe-K} absorption feature) is
significantly different from the two-component spectrum we presented here. This
strongly suggests that {\it this kind of short-term variability is related not to a
specific energy spectral component, but to the nature that these systems
can produce superluminal jets.} This variability is, however, not directly
related to the
presence of jets no significant radio emission during these
observations. 

\section{Summary}

The broad band high energy observations presented here have for the first time
detected a two-component energy spectrum and short-term variability in the 1-10
keV band from a superluminal jet source. The former is shared by all types of
identified BHBs, but not by any known neutron star system, and is therefore a
firm signature from BHBs. The latter is unusual among BHBs, and is probably
somehow related to uniqueness of this source as a superluminal jet BHB. We have
constrained the mass of the central compact object and the binary system
inclination angle using the X-ray energy spectrum combined with optical radial
velocity measurements. Our results further support the system containing a
black hole with the highest inclination angle of all known BHBs in our Galaxy.
The origin of the hard X-ray power-law component is still uncertain.
Thermal Comptonization is unlikely to be the hard X-ray production mechanism,
in view of
the rather steep and unbroken power-law up to very high energies. Jet origin
models also seem difficult due to the apparent lack of significant radio
emission during this and the previous hard X-ray outburst. The converging flow
model seems consistent with the spectral data, but requires the hard X-ray
production region to be compact. This can be tested if this system is an X-ray
eclipsing BHB. The origin of the short-term variability in the 1-10 keV band is
also puzzling. This is perhaps related to the unique nature of this source as a
superluminal jet BHB. 

\section{Acknowledgement}

We appreciate the supports of the ASCA, BATSE and WATCH teams for data 
collection and analysis. We also thank Jan van Paradijs and Craig Robinson
for many interesting discussions. Finally we are very grateful to the referee 
R.M. Hjellming for comments and suggestions, which certainly improved this
article.

{\it Note added in proof.} After the acceptance of this article for
publication, we became aware of new results obtained by Orosz and Bailyn (1997)
(ApJ, in press) from optical light curve modeling of GRO~J1655-40. Their
mass value (7.02 $\pm$ 0.22 M$_{\sun}$) and the binary inclination angle (69.50
$\pm$ 0.08 degrees) are significantly different from what we obtained by
combining the X-ray energy spectrum and the optical mass function. 
The apparent discrepancy might be
caused by any of several factors. For example, if the black hole is rotating in
the same direction as the
accretion disk, the curves in our figure 4 would all move upward, implying
a higher black hole mass for a given inclination angle. Applying the
inferred black hole mass value and
the inclination angle from Orosz and Bailyn to our X-ray {\it M-i} 
relationship and assuming
the black hole has non-zero angular momentum ({\it Kerr} black hole), we obtain
an inner disk radius of $\sim$ 2{\it GM/c}$^{2}$, to be compared with 6{\it
GM/c}$^{2}$ for a {\it Schwarzchild} (non-rotating) black hole. This inner disk
radius would correspond to the last stable orbit of a $\sim$95\% maximally
rotating black hole. Also, the X-ray
spectral model we used may not be sufficiently accurate when the inclination
angle is very high and the inner disk boundary is very close to the horizon of
the black hole.

\newpage
\figcaption[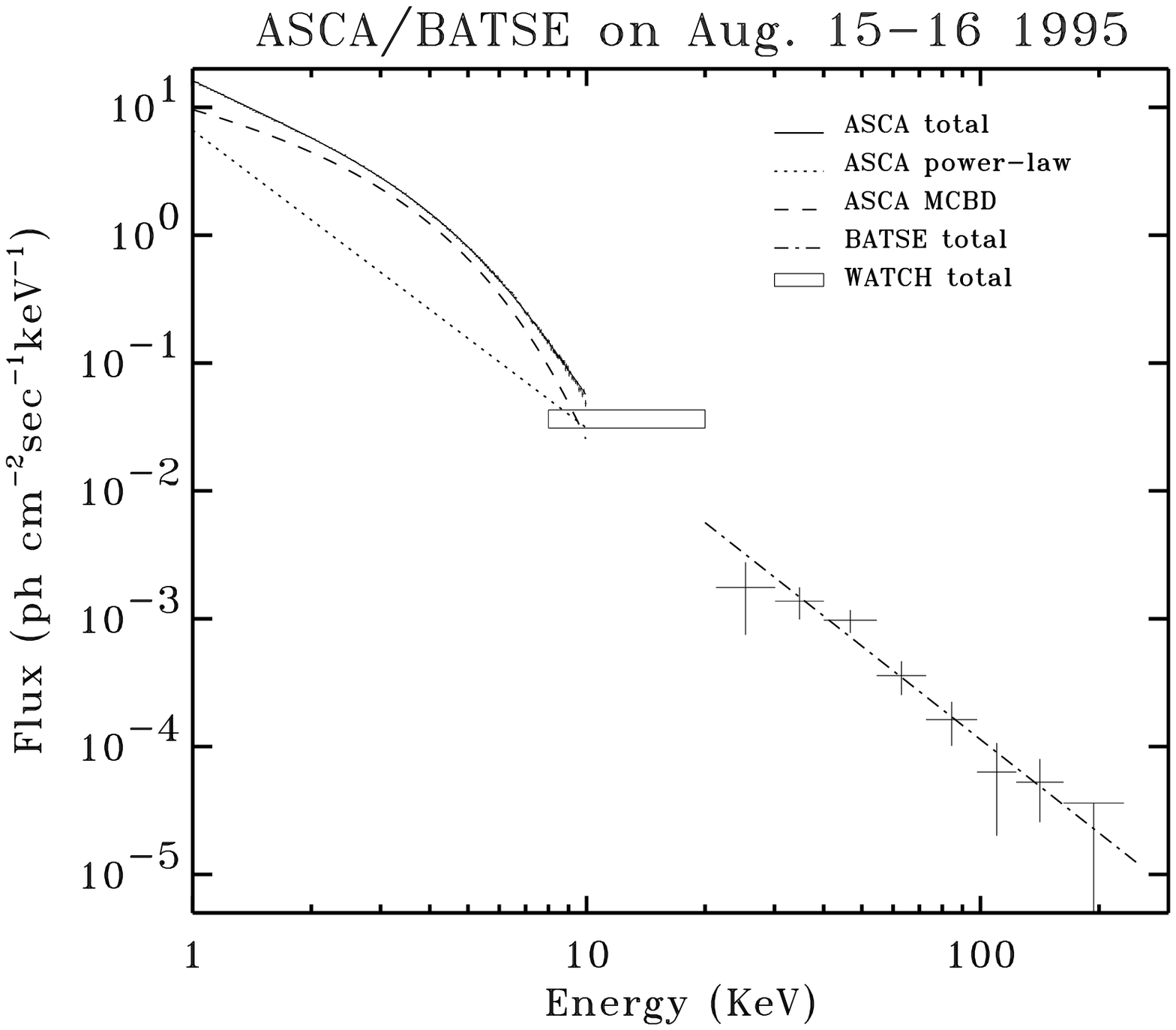]{Joint ASCA/BATSE spectrum with WATCH integrated flux. Spectral models and data points are
shown (ASCA error bars are extremely small). The total ASCA energy spectrum is
composed of two components, i.e., an ultra-soft component described by
a multi-color blackbody disk (MCBD) model and a power-law component which is 
consistent,
when extrapolated into the BATSE energy range,
in both photon index and flux with the independently determined BATSE 
power-law. The WATCH data point is consistent with the sum of these two 
components. No third spectral component is required for the entire 1-100 keV
spectrum, except for a weak absorption-line feature in the iron K-fluorescence
energy band of the ASCA data. \label{figure1}}

\figcaption[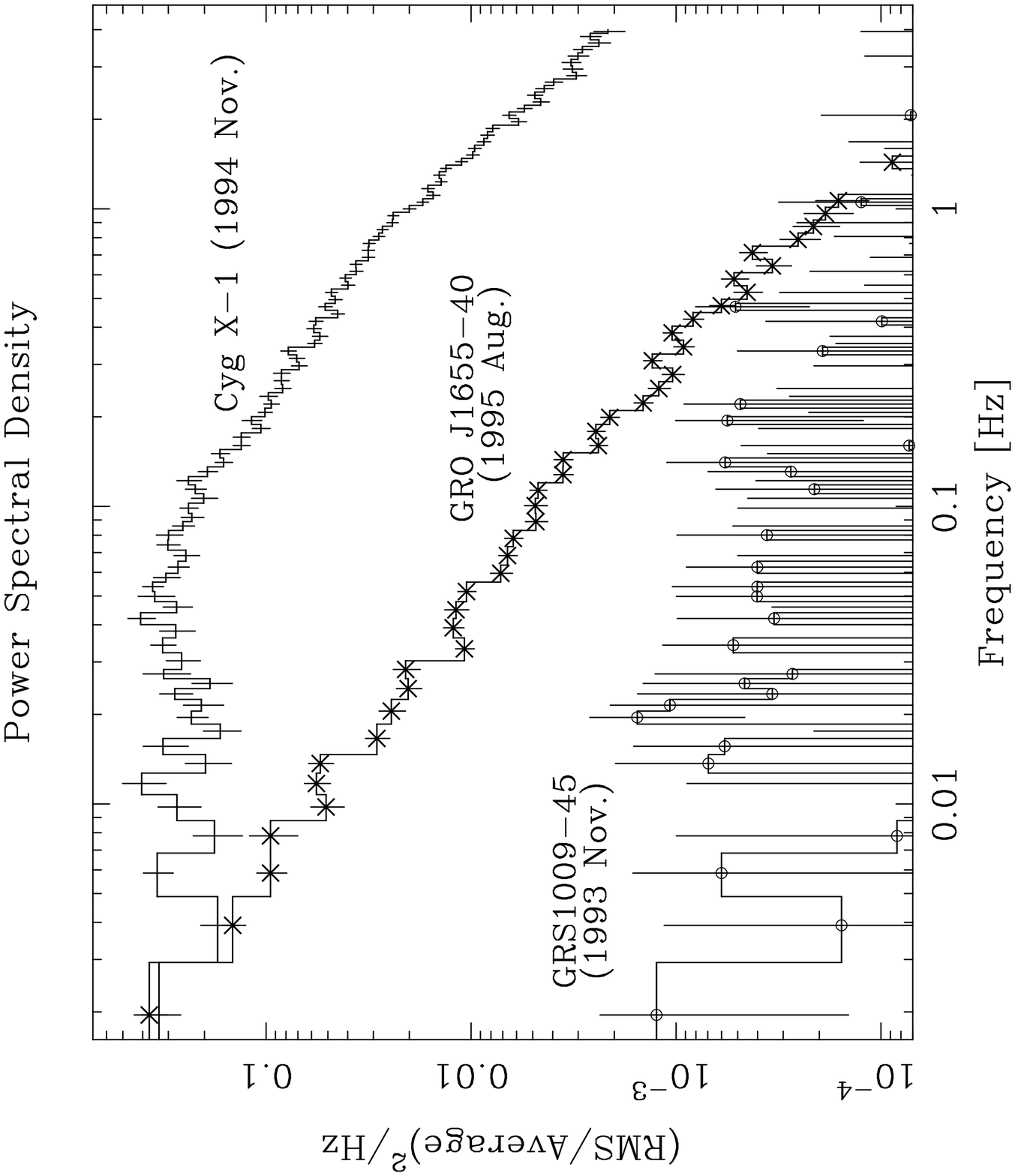]{Normalized power density spectra of GRO~J1655--40, Cygnus X-1 and 
GRS~1009--45, obtained with the ASCA GIS monitor. The two BHCs
Cygnus~X-1 and GRS~1009-45 were in a typical low state and high state, 
respectively. The power-law PDS of GRO~J1655-40 is very unusual among BHBs.
\label{figure2}}

\figcaption[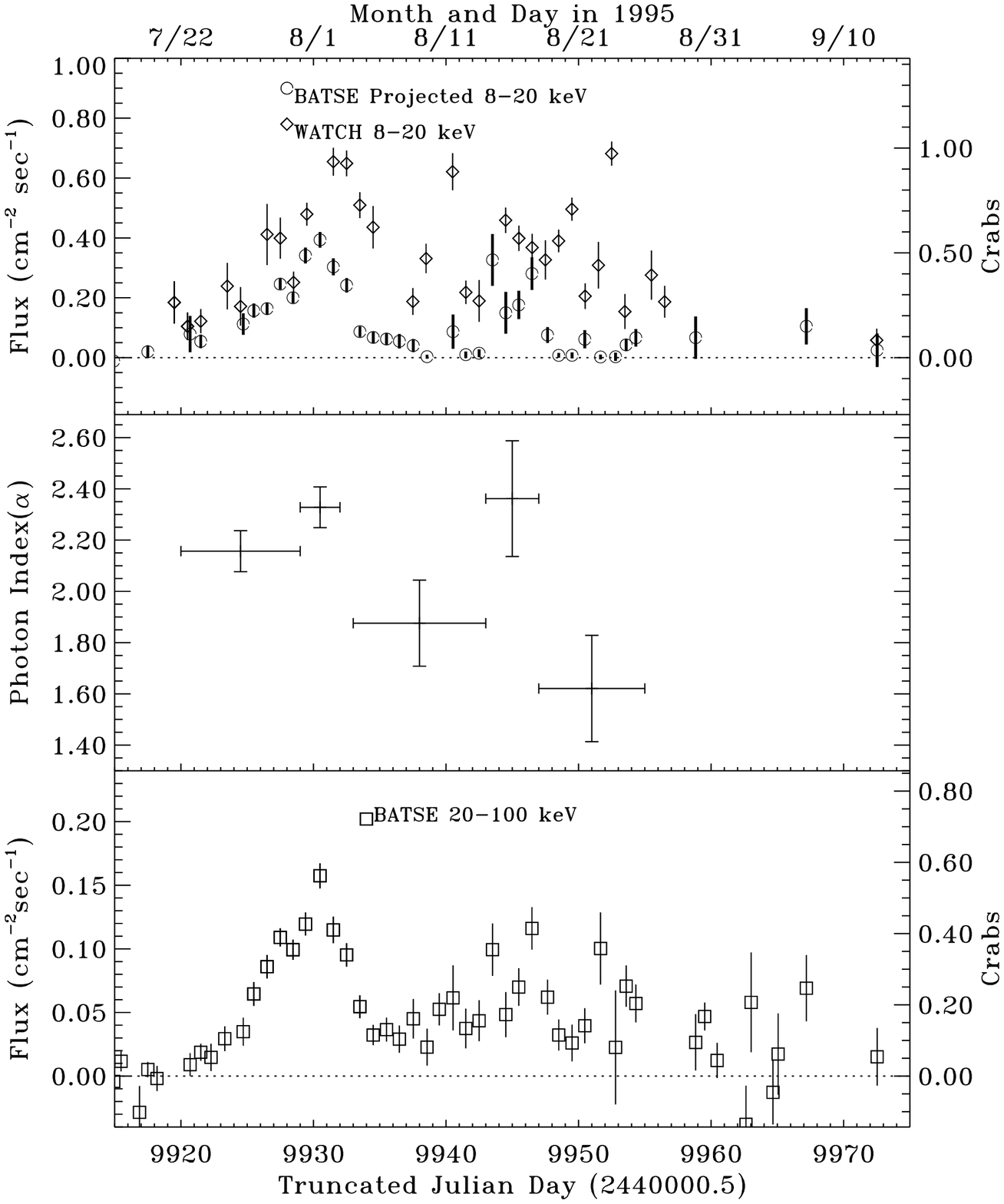]{BATSE and WATCH light curves and spectral variations. The top panel
displays the WATCH 8-20 keV light curve (diamonds) and the projected
flux in 8-20 keV from power-law fits to the 20-100 keV BATSE count spectra (circles). 
The spectral index was allowed to vary. The total 8-20
keV fluxes detected by WATCH are always above the 20-100 keV power-law
extrapolation, indicating the existence of an additional soft component. The
middle panel shows the photon spectral index of the 20-100 keV power-law
observed by BATSE. The bottom panel is the BATSE light curve in the
20-100 keV band. \label{figure3}}

\figcaption[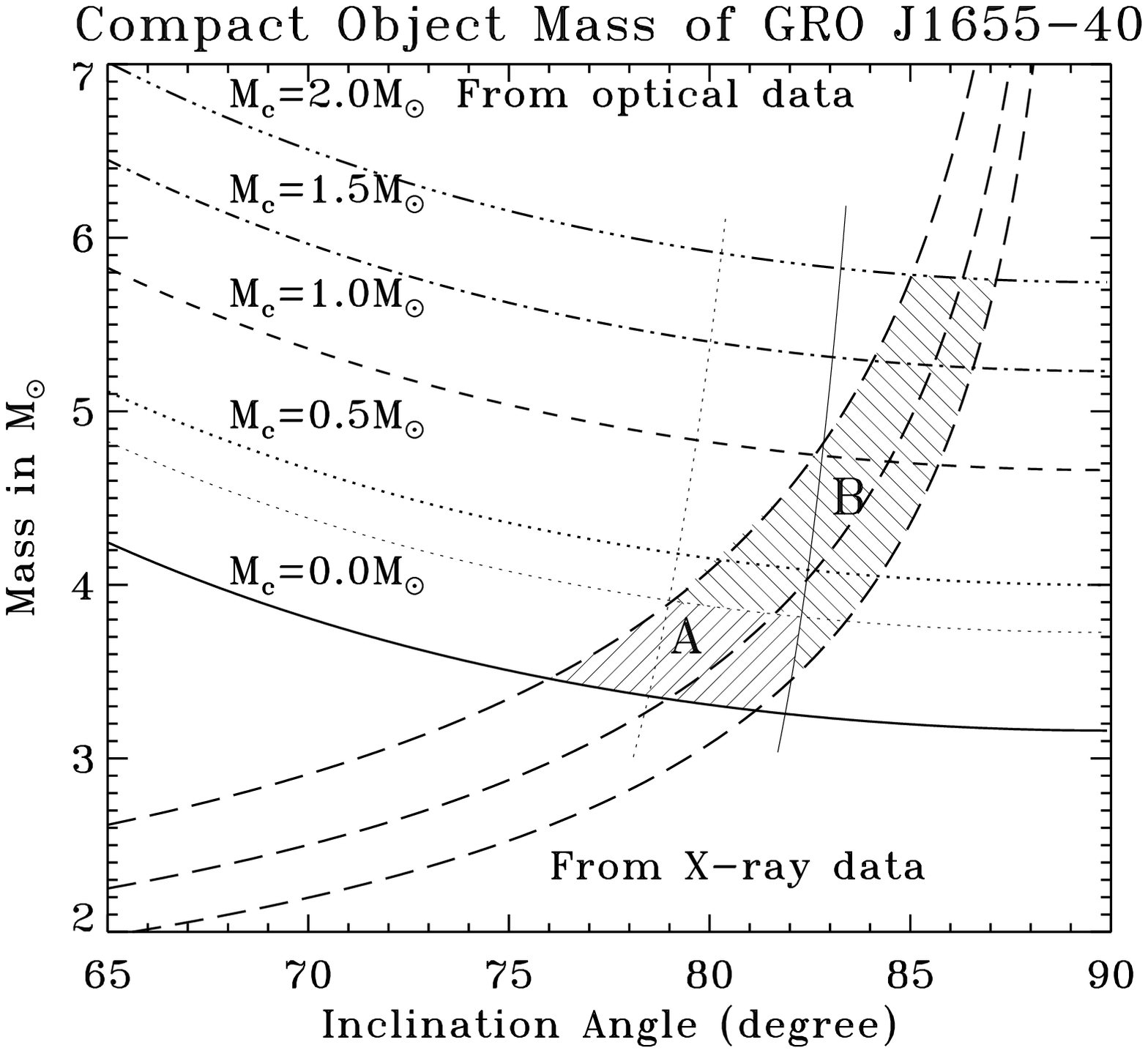]{The relationships between the mass of the central compact object
(black hole) and the binary system inclination angle, as determined from
optical radial velocity and X-ray spectroscopy measurements. The five nearly
parallel and horizontal (thick) lines are from the optical radial velocity
determination of the mass function (Bailyn, \etal\  1995b), corresponding to the
companion mass of 0.0, 0.5, 1.0, 1.5 and 2.0 M$_{\sun}$. The three nearly
parallel curves from the lower left corner to the upper right corner are from
the inner disk radius determined from X-ray spectroscopy. The allowed parameter
space is indicated by two shaded areas, { `A'} (non-eclipsing zone) and {
`B'} (eclipsing zone) respectively (see text for details). The range of the
compact object mass is constrained to be from 3.3 to 5.8 M$_{\sun}$. The compact
object is thus likely a black hole. The inclination angle of the binary system
is between 76 and 87 degrees. 
\label{figure4}}

\newpage
\setcounter{figure}{0}
\begin{figure}
\centerline{
\psfig{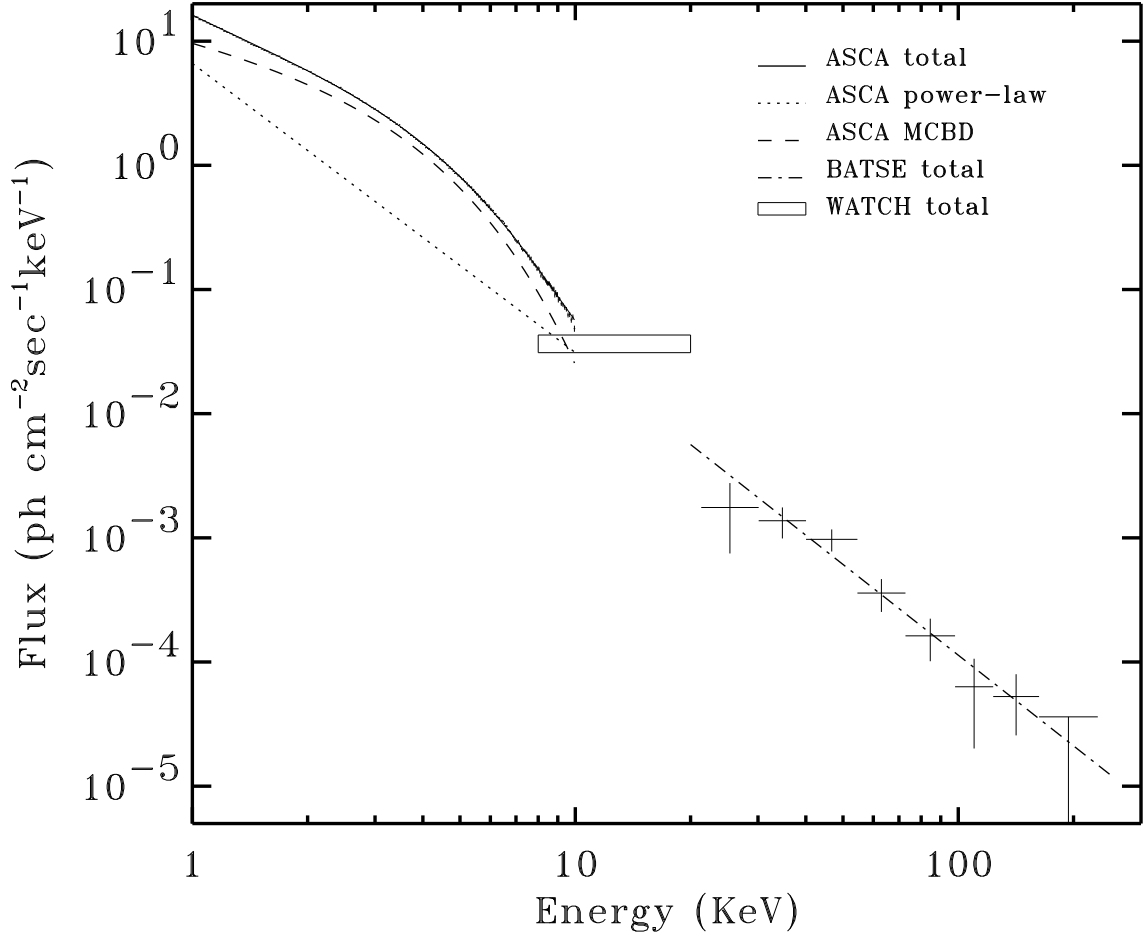}
}
\caption{Joint ASCA/BATSE spectrum with WATCH integrated flux. Spectral models and data points are
shown (ASCA error bars are extremely small). The total ASCA energy spectrum is
composed of two components, i.e., an ultra-soft component described by
a multi-color blackbody disk (MCBD) model and a power-law component which is 
consistent,
when extrapolated into the BATSE energy range,
in both photon index and flux with the independently determined BATSE 
power-law. The WATCH data point is consistent with the sum of these two 
components. No third spectral component is required for the entire 1-100 keV
spectrum, except for a weak absorption-line feature in the iron K-fluorescence
energy band of the ASCA data.}
\end{figure}

\begin{figure}
\centerline{
\rotate[r]{
\psfig{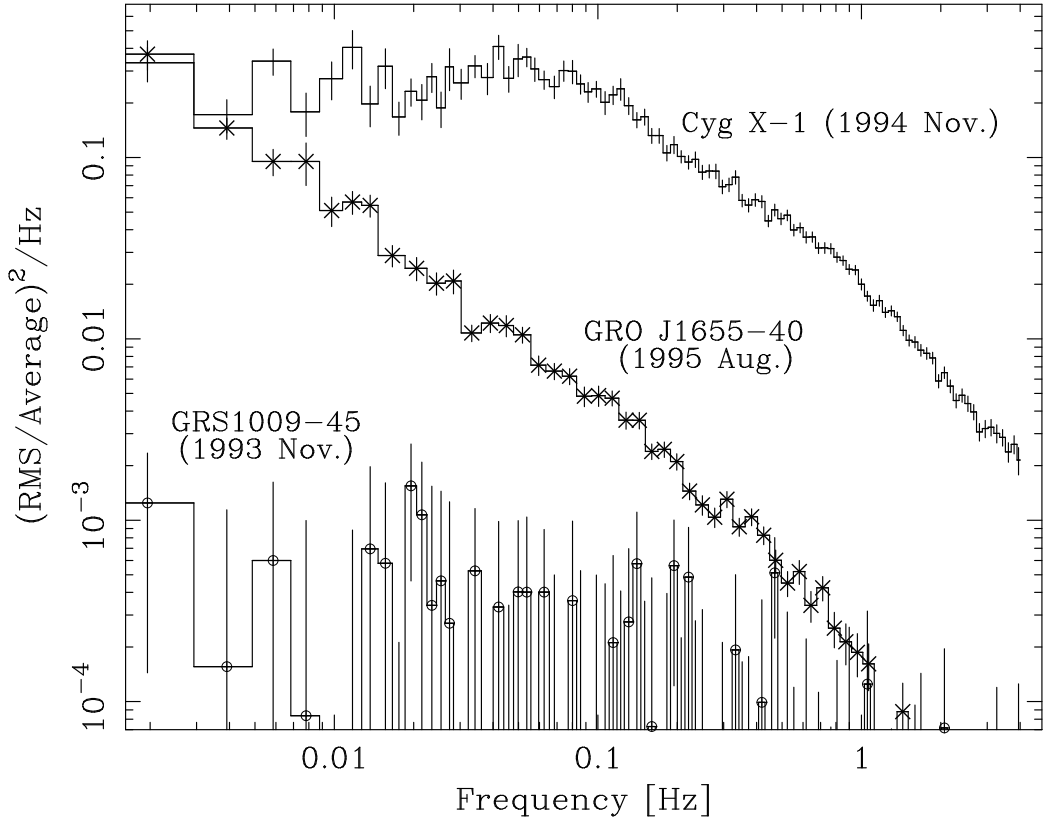}
}
}
\caption{Normalized power density spectra of GRO~J1655--40, Cygnus X-1 and 
GRS~1009--45, obtained with the ASCA GIS monitor. The two BHCs
Cygnus~X-1 and GRS~1009-45 were in a typical low state and high state, 
respectively. The power-law PDS of GRO~J1655-40 is very unusual among BHBs.
}
\end{figure}

\begin{figure}
\rotate[l]{
\begin{minipage}[b]{5.0in}
\centerline{
\psfig{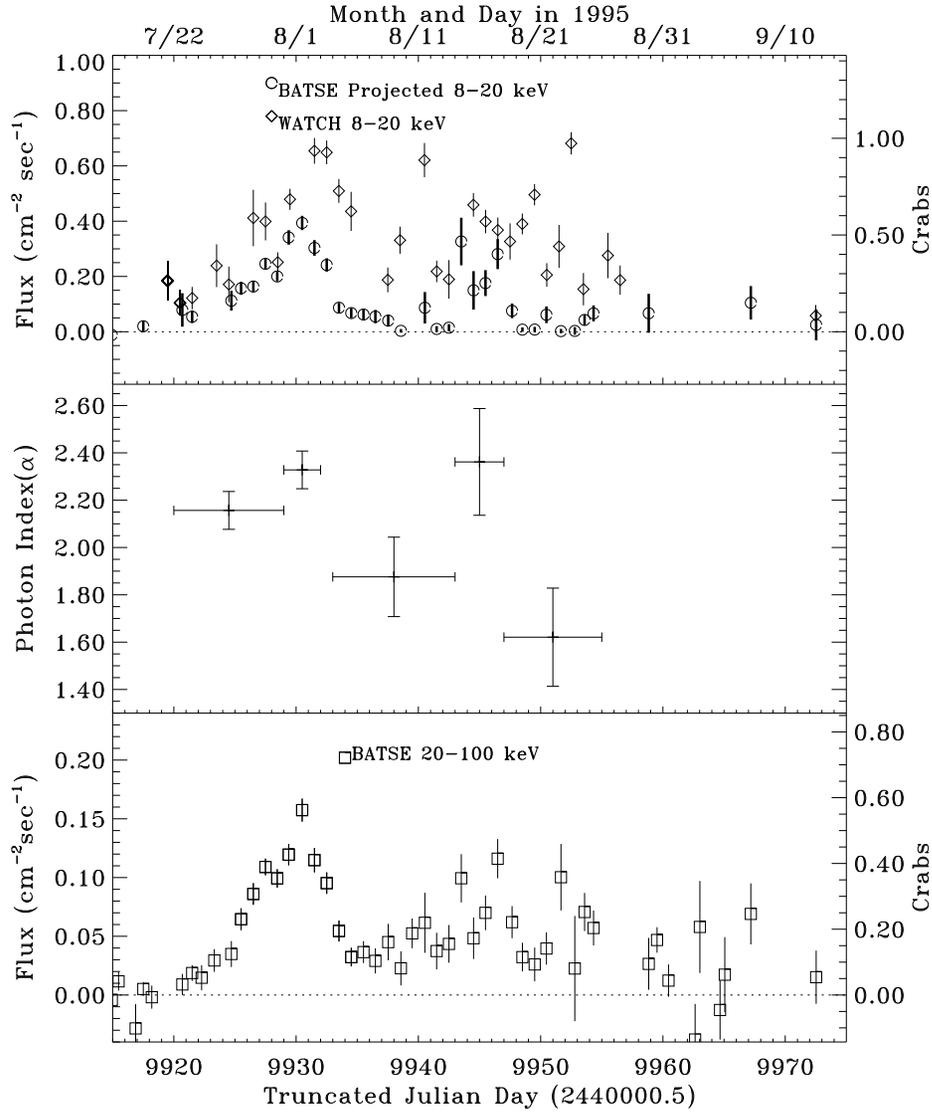}
}
\end{minipage}\hfill
\begin{minipage}[b]{3.5in}
\caption{BATSE and WATCH light curves and spectral variations. The top panel
displays the WATCH 8-20 keV light curve (diamonds) and the projected
flux in 8-20 keV from power-law fits to the 20-100 keV BATSE count spectra (circles). 
The spectral index was allowed to vary. The total 8-20
keV fluxes detected by WATCH are always above the 20-100 keV power-law
extrapolation, indicating the existence of an additional soft component. The
middle panel shows the photon spectral index of the 20-100 keV power-law
observed by BATSE. The bottom panel is the BATSE light curve in the
20-100 keV band.} 
\end{minipage}\hfill
}
\end{figure}

\begin{figure}
\centerline{
\psfig{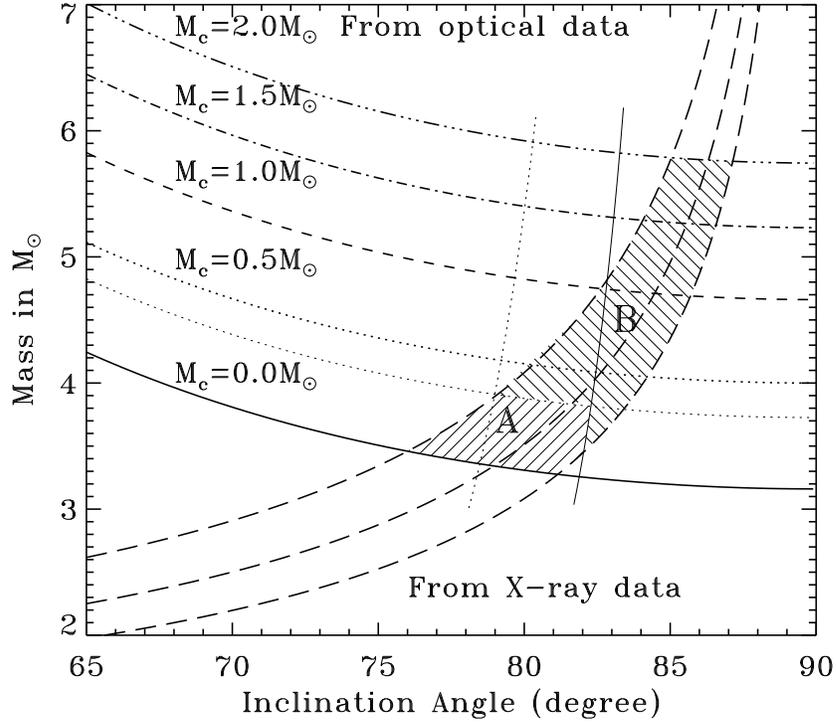}
}
\caption{The relationships between the mass of the central compact object
(black hole) and the binary system inclination angle, as determined from
optical radial velocity and X-ray spectroscopy measurements. The five nearly
parallel and horizontal (thick) lines are from the optical radial velocity
determination of the mass function (Bailyn, \etal\  1995b), corresponding to the
companion mass of 0.0, 0.5, 1.0, 1.5 and 2.0 M$_{\sun}$. The three nearly
parallel curves from the lower left corner to the upper right corner are from
the inner disk radius determined from X-ray spectroscopy. The allowed parameter
space is indicated by two shaded areas, { `A'} (non-eclipsing zone) and {
`B'} (eclipsing zone) respectively (see text for details). The range of the
compact object mass is constrained to be from 3.3 to 5.8 M$_{\sun}$. The compact
object is thus likely a black hole. The inclination angle of the binary system
is between 76 and 87 degrees. 
}
\end{figure}
\end{document}